%% file: ms.tex
\documentclass{aastex}
\usepackage{emulateapj5,times,mathptm}
\journalinfo{The Astrophysical Journal Letters, 2002 December, astro-ph/0210552}
\slugcomment{Received 2002 October 9; accepted 2002 October 24}

\shorttitle{$z>4$ AGNs in the 2~Ms \cdfn\ Survey}
\shortauthors{VIGNALI ET AL.}

\newcommand{\ltsima}{$\; \buildrel < \over \sim \;$}
\newcommand{\simlt}{\lower.5ex\hbox{\ltsima}}
\newcommand{\gtsima}{$\; \buildrel > \over \sim \;$}
\newcommand{\simgt}{\lower.5ex\hbox{\gtsima}}

\newcommand{\cgs}{ ${\rm erg~cm}^{-2}~{\rm s}^{-1}$ }

\def\sun{\hbox{$\odot$}}
\def\earth{\hbox{$\oplus$}}
\def\lesssim{\mathrel{\hbox{\rlap{\hbox{\lower4pt\hbox{$\sim$}}}\hbox{$<$}}}}
\def\gtrsim{\mathrel{\hbox{\rlap{\hbox{\lower4pt\hbox{$\sim$}}}\hbox{$>$}}}}

\def\arcdeg{\hbox{$^\circ$}}
\def\arcmin{\hbox{$^\prime$}}

\def\aox{$\alpha_{\rm ox}$}

\def\lumh{\rm erg s$^{-1}$ Hz$^{-1}$}
\def\ab1450{$AB_{1450(1+z)}$}
\def\one{CXOHDFN~J123647.9$+$620941}
\def\two{CXOHDFN~J123719.0$+$621026}
\def\three{CXOHDFN~J123642.0$+$621331}
\def\oner{CXOHDFN~J123647.9$+$620941}
\def\twor{CXOHDFN~J123719.0$+$621026}
\def\threer{CXOHDFN~J123642.0$+$621331}
\def\cdfn{\hbox{CDF-N}}
\def\hdfn{\hbox{HDF-N}}
\def\mb{$M_{\rm B}$~}

\def\chandra{{\it Chandra\/}}

\def\heao1{{\it HEAO-1\/}}

\def\rosat{{\it ROSAT\/}}

\def\xmm{{XMM-{\it Newton\/}}}

\begin{document}

\title{The Chandra Deep Field-North Survey. XVI. 
The X-ray properties of moderate-luminosity Active Galaxies at $z>4$}
\author{
C. Vignali,\altaffilmark{1} 
F.~E. Bauer,\altaffilmark{1} 
D.~M. Alexander,\altaffilmark{1} 
W.~N. Brandt,\altaffilmark{1} 
A.~E. Hornschemeier,\altaffilmark{2}
D.~P. Schneider,\altaffilmark{1}
and G.~P. Garmire\altaffilmark{1} 
}

\altaffiltext{1}{Department of Astronomy \& Astrophysics, The Pennsylvania State University, 
525 Davey Laboratory, University Park, PA 16802, USA \\
({\tt chris, fbauer, davo, niel, dps, and garmire@astro.psu.edu}).}
\altaffiltext{2}{Chandra Fellow, Department of Physics \& Astronomy, 
Johns Hopkins University, 3400 N. Charles Street, Baltimore, MD 21218, USA \\
({\tt annh@pha.jhu.edu}).}


\begin{abstract}

We present X-ray spectral analyses of the three $z>4$ Active Galactic Nuclei (AGNs) 
thus far spectroscopically identified in the \chandra\ Deep Field-North Survey, at redshifts of 5.186, 4.424, and 4.137. 
These analyses are made possible by the extremely deep exposure ($\approx$~2~Ms) and the low \chandra\ background. 
The rest-frame \hbox{$\approx$~2.5--40~keV} spectra are the first for optically faint 
(two of the three sources have $I>24$) $z>4$ AGNs. 
The $z=5.186$ quasar is well fitted by a power-law model with photon 
index $\Gamma=1.8\pm{0.3}$, consistent with those of lower-redshift, unobscured AGNs. 
The other two AGNs have flatter effective \hbox{X-ray} photon indices \hbox{($\Gamma\approx$~1.1--1.5)}, 
suggesting the presence of intrinsic absorption (provided their underlying \hbox{X-ray} continua 
are similar to those of lower-redshift AGNs). 
It is possible that the flat \hbox{X-ray} continuum of the $z=4.424$ AGN is partially related to its radio loudness. 
If the $z=4.137$ AGN suffers from \hbox{X-ray} absorption, the implied column density is 
\hbox{$N_{\rm H}\approx2\times10^{23}$~cm$^{-2}$}. 

\end{abstract}


\section{Introduction}

One of the most challenging goals in modern astronomy is to investigate the nature of the first objects to form 
at the end of the ``Dark Age'' (i.e., at the reionization epoch; e.g., Rees 1999; Loeb \& Barkana 2001). 
From an X-ray perspective, 
there is increasing evidence that Active Galactic Nuclei (AGNs) are powerful X-ray emitters in the early Universe 
(e.g., at $z>4$; Kaspi, Brandt, \& Schneider 2000; Vignali et al. 2001, 2003, hereafter V01 and V03; Brandt et al. 2002a,b) 
as well as locally. 
However, to date \hbox{X-ray} spectral analysis at $z>4$ has only been possible 
for the most optically luminous quasars (e.g., \mb$\approx$~$-$27 to $-$30; V03), 
which are presumably only a minority of the AGN population. 

In this study we use the 2~Ms exposure of the 
\chandra\ Deep Field-North (\cdfn) to constrain the X-ray properties of three moderate-luminosity $z>4$ AGNs. 
This is the first direct \hbox{X-ray} spectral analysis performed on $z>4$ radio-quiet AGNs 
with more than 100 counts (two of the three sources). 
Indeed, constraints on the \hbox{X-ray} properties of moderate-luminosity AGNs 
at the highest redshifts via \hbox{X-ray} spectral analysis 
can only be placed by deep \hbox{X-ray} surveys at present. 
The moderate-luminosity and presumably moderate-mass AGNs found in the \cdfn\ at $z>4$ 
are likely related to the smallest cold dark matter haloes. 
In hierarchical large-scale structure formation, these small haloes collapse earlier than the larger ones, 
thus effectively tracing the first massive structures to form in the Universe. 

The Galactic column density along the line of sight to these sources is $1.6\times10^{20}$~cm$^{-2}$ 
(Stark et al. 1992). 
In this Letter $H_{0}=70$~km~s$^{-1}$~Mpc$^{-1}$, $\Omega_{\rm M}$=0.3, and $\Omega_{\Lambda}$=0.7 
are adopted.

\section{X-ray observations and the sample of $z>4$ AGNs}

The X-ray results were obtained with ACIS-I (the imaging array of the Advanced CCD 
Imaging Spectrometer; Garmire et al. 2002) onboard \chandra. 
The 2~Ms \cdfn\ observations were centered on the Hubble Deep Field-North (\hdfn; Williams et al. 1996) 
and cover \hbox{$\approx$~460~arcmin$^2$}. They reach on-axis \hbox{0.5--2.0~keV} 
(soft-band) and \hbox{2--8~keV} (hard-band) flux limits of \hbox{$\approx1.5\times10^{-17}$~\cgs} and 
\hbox{$\approx1.0\times10^{-16}$~\cgs}, respectively. 
The 2~Ms data processing and the derived X-ray catalog are described in D.~M. Alexander et al., in preparation. 

We investigate here the \hbox{X-ray} properties of the three \hbox{$z>4$} AGNs 
spectroscopically identified in the \cdfn: 
the \hbox{$z=5.186$} quasar \one\ (Barger et al. 2002, hereafter B02; \mb$\approx-23.4$), and the two Seyfert galaxies 
\two\ (B02; \mb$\approx-21.4$) and \three\ (VLA~J123642$+$621331; Waddington et al. 1999, hereafter W99; Brandt et al. 2001b; 
\hbox{\mb$\approx-21.6$}) 
at $z=4.137$ and $z=4.424$, respectively; the latter Seyfert galaxy 
is the only object in our sample with detected radio emission (0.47~mJy at 1.4~GHz; Richards 2000). 
%
%
Only one object (\two) shows hints of \hbox{X-ray} variability by a factor of $\approx$~2 
over the \hbox{$\approx$~27-month} interval covered by the 20 \hbox{X-ray} observations that 
comprise the 2~Ms exposure (F.~E. Bauer et al., in preparation). 
We note, however, that our \hbox{X-ray} constraints 
are weak because of the low counting statistics in each observation.  
A summary of the principal optical and \hbox{X-ray} properties of the 
three AGNs is presented in Table~1. 

We would expect there to be further unidentified $z>4$ AGNs in the \cdfn.
For instance, using the surface density of $z>4$ quasars in the Sloan Digital Sky Survey (SDSS) Early Data Release 
quasar catalog (Schneider et al. 2002) and the luminosity function derived from $z\approx$~4.3 
SDSS quasars (Fan et al. 2001b), 
we would expect $\approx$~7 (0.5) AGNs over the \cdfn\ field down to \mb=$-$21.4 ($-$23.4). 
This is likely to be a lower limit since this 
calculation does not take into account narrow-lined AGNs. 
%
\input{tabb1.tex}
%
All of the sources within \hbox{$\approx$~10\arcmin--12\arcmin} of the aimpoint should be detectable at 
X-ray energies [assuming \aox=$-$1.48 (see $\S$4), 
we predict soft \hbox{(0.5--2~keV)} \hbox{X-ray} fluxes of \hbox{$\approx$(6--8)$\times10^{-17}$~\cgs}]. 
However, with predicted optical magnitudes down to $I\approx$~25, many will be optically faint and challenging to
identify (e.g., Alexander et al. 2001).

\section{X-ray spectral analysis}

Given the deep exposure and the low (but non-negligible) background 
of the ACIS instrument, we are able to perform direct spectral analyses for the sources in our sample.  
Since the 20 separate observations that comprise the 2~Ms \cdfn\ 
are characterized by different roll angles and aimpoints, we used the 
source extraction code ({\sc ACIS Extract}\footnote{{\sc ACIS Extract} is a part 
of the {\sc TARA} software package and can be accessed from 
http://www.astro.psu.edu/xray/docs/TARA/ae\_users\_guide.html.}) 
described in P.~S. Broos et al. (2002). 
%
For each source, this code extracts the counts from 
each of the 20 observations, taking into account the dependence in shape and size of the Point Spread Function (PSF) 
with off-axis angle as given in 
the \chandra\ \hbox{X-ray} Center (CXC) PSF library.\footnote{See http://asc.harvard.edu/ciao2.2/documents\_dictionary.html\#psf.} 
For the sources investigated here, events are extracted using the 90\% encircled energy regions measured at 1.5~keV 
(at the position of each source). 
The background was chosen locally after all sources were excluded from the \hbox{X-ray} event file. 
Visual inspection of the \hbox{X-ray} images shows that the three AGNs are not contaminated by 
nearby \hbox{X-ray} sources. 
To account for the quantum efficiency degradation of 
ACIS\footnote{See http://cxc.harvard.edu/cal/Links/Acis/acis/Cal\_prods/qeDeg/index.html.} 
at low energies ($\approx$~10\% at 1~keV), possibly caused by  
molecular contamination of the ACIS filters, we have applied a time-dependent correction to the ACIS 
quantum efficiency generating a corrected ancillary response file (ARF) 
for each data set (G. Chartas et al., in preparation).\footnote{See http://www.astro.psu.edu/users/chartas/xcontdir/xcont.html.} 
An energy-dependent aperture correction was also applied to each ARF. 
Spectra and response matrices, weighted by the number of counts in each observation, 
were summed using standard {\sc FTOOLS} routines (Blackburn 1995). 

Spectral analysis was carried out with {\sc XSPEC} (Version 11.2.0; Arnaud 1996); 
the uncertainties on spectral parameters 
are quoted at the 90\% confidence level for one interesting parameter (i.e., $\Delta\chi^{2}$=2.71; Avni 1976). 
Given the limited counting statistics, 
%
%
we performed the spectral analysis using the unbinned, background-subtracted 
source spectra and the Cash statistic (Cash 1979). 
The Cash statistic is well suited to low-count sources (e.g.,\ Nousek \& Shue 1989), and in the latest version of 
{\sc XSPEC} it is possible to use it with background-subtracted data (Arnaud 2002). 
Several checks of the background-subtraction method were carried out to verify that 
no spurious residual features were present in the background-subtracted data. 
One limitation of the Cash statistic is that it does not provide a quality-of-fit criterion 
(like the $\chi^{2}$ statistic) to compare different models. 
Therefore, we established the quality of the spectral results via visual inspection 
using the binned spectrum. 
In the spectral fitting we have assumed solar abundances, although there are indications of 
supersolar abundances of heavy elements in high-redshift quasar nuclei 
(e.g., Hamann \& Ferland 1999; Constantin et al. 2002). 
Choosing different abundances provides changes in the column density 
and Fe~K$\alpha$ line measurements. For example, 
doubling the abundances in the fit gives a reduction in the column density by $\approx$~50\% 
and a small increase (a few \%) in the iron~K$\alpha$ line intensity (see Fig.~17 of George \& Fabian 1991).

\subsection{Spectral analysis results}

\oner\ is the only $z>5$ quasar 
having an \hbox{X-ray} spectrum with more than 100 counts to date. 
Spectral fitting in the observed-frame 
\hbox{0.5--8~keV} band (Fig.~1) shows that 
this source is well parameterized by a power-law continuum with photon index \hbox{$\Gamma=1.81^{+0.31}_{-0.29}$} 
(see Table~2). This photon index is 
similar to those of $z\approx$~0--2 AGNs, 
which have typical photon indices of $\Gamma\approx$~1.7--2.3 in the rest-frame \hbox{2--10~keV} band 
(e.g., George et al. 2000; Reeves \& Turner 2000). 
%
\figurenum{1}
\centerline{\includegraphics[height=8.5cm,angle=-90]{f1.ps}}
\figcaption{\footnotesize 
\oner\ spectrum (binned to 10 counts per bin for presentation purposes) 
with the best-fit $\Gamma=1.81$ power law and Galactic absorption. 
Data-to-model residuals are shown in the bottom panel (in units of $\sigma$).  
The dashed vertical line indicates the observed-frame energy for 
a neutral Fe~K$\alpha$ emission line. 
\label{fig1}}
\centerline{}
%
\input{tabb2.tex}
%
There is no evidence for either intrinsic absorption or Fe~K$\alpha$ emission lines (see Table~2).  
The optical spectrum of \oner\ shows a Ly$\alpha$ emission line (B02) 
whose FWHM appears significantly smaller than those typical of optically selected, optically bright, high-redshift quasars 
(e.g., Fan et al. 2001a). The lack of flux calibration in the spectrum of B02 
prevents a more precise comparison. 

\twor\ is characterized by a flat effective \hbox{X-ray} photon index ($\Gamma=1.12\pm{0.25}$), which provides 
an acceptable fit to our data. There are two possible explanations for this result, 
both consistent with our data: \twor\ has either an intrinsically flat \hbox{X-ray} spectrum or 
a steeper ($\Gamma\approx$~2.0) \hbox{X-ray} spectrum plus absorption. 
Based on previous results for low- and intermediate-redshift quasars (e.g., George et al. 2000; 
Reeves \& Turner 2000), optically luminous $z>4$ quasars (V03), and \oner, we suggest 
that the underlying continuum is likely a $\Gamma\approx$~2.0 power law. 
When the photon index is fixed to 2, a deficit of \hbox{X-ray} counts is visible below 
$\approx$~3~keV (see Fig.~2). 
%
\figurenum{2}
\centerline{\includegraphics[height=8.5cm,angle=-90]{f2.ps}}
\figcaption{\footnotesize 
\twor\ spectrum (binned to 10 counts per bin for presentation purposes) 
fitted above 3~keV with a $\Gamma=2$ power-law model that has been extrapolated back to lower energies. 
Data-to-model residuals are shown in the bottom panel (in units of $\sigma$).  
A deficit of counts below $\approx$~3~keV is present. 
The dashed vertical line indicates the observed-frame energy for 
a neutral Fe~K$\alpha$ emission line. 
\label{fig2}}
\vglue0.2cm 
%
This suggests the presence of a column density of \hbox{$\approx$~2$\times10^{23}$~cm$^{-2}$} (see Table~2), 
typical of local Compton-thin Seyfert~2 galaxies (e.g., Bassani et al. 1999). 
The upper limit on the neutral Fe~K$\alpha$ emission-line equivalent width (rest-frame $EW<660$~eV) 
allows us to rule out the possibility that the \hbox{X-ray} emission is due purely to a 
scattered/reflected component from a Compton-thick source (i.e., a source with a column density larger than 
$1.5\times10^{24}$~cm$^{-2}$; e.g., Matt et al. 2000), since in this case a strong ($EW\simgt1$~keV) iron K$\alpha$ 
line at 6.4~keV is expected (e.g., Matt, Brandt, \& Fabian 1996; Maiolino et al. 1998). 
The flat \aox\ value (see Table~2 and $\S$4) suggests that the AGN is comparitively weak at optical wavelengths; 
the presence of a narrow Ly$\alpha$ line supports this suggestion.

\threer\ is the faintest \hbox{X-ray} source in our sample and lies just outside the \hdfn. 
Detailed studies of this source indicate a redshift of $z=4.424$ 
(W99; R.~A. Windhorst 2001, private communication; 
but also see $\S$2.2 of Barger, Cowie, \& Richards 2000). 
The presence of an AGN is suggested by multi-wavelength studies. 
The apparent radio jet in the combined MERLIN/VLA image (W99) indicates the presence of an AGN, 
as does the source's large 1.4~GHz radio luminosity \hbox{($\approx7.8\times10^{32}$~\lumh)}, 
which is more typical of AGNs than starburst galaxies (e.g., Bauer et al. 2002). 
From an \hbox{X-ray} perspective, 
Brandt et al. (2001b), based on a 480~ks \chandra\ observation, 
argued that its \hbox{X-ray}-to-optical flux ratio and \hbox{X-ray} luminosity were 
suggestive of the presence of an active nucleus. 
This is confirmed by the 2~Ms results reported in this Letter. In particular, 
the detection of \threer\ in the \hbox{2--8~keV} band (corresponding to the rest-frame \hbox{$\approx$~11--43~keV band}) 
strongly supports the presence of an active nucleus rather than only a starburst. 
The relatively flat \hbox{X-ray} spectrum \hbox{($\Gamma=1.55^{+0.61}_{-0.54}$)} can be due either to the presence of 
intrinsic absorption (not constrained by the present observation; see Table~2) 
or to the radio loudness of this source.\footnote{Its radio loudness, parameterized by 
$R$ = $f_{\rm 5~GHz}/f_{\rm 4400~\mbox{\scriptsize\AA}}$ (rest frame; e.g., Kellermann et al. 1989), is $\approx$~1200.} 
In fact, previous studies (e.g., Wilkes \& Elvis 1987; Cappi et al. 1997) indicate that  
jet emission can give rise to flatter slopes in the \hbox{X-ray} band because of 
synchrotron or synchrotron self-Compton emission.

\section{Broad-band comparisons with other $z\ge4$ AGNs}

Presently, $\approx$~50 $z\ge4$ quasars have \hbox{X-ray} detections (see Brandt et al. 2002b and V03 for the most recent 
results).\footnote{Also see http://www.astro.psu.edu/users/niel/papers/highz-xray-detected.dat for a 
regularly updated compilation of X-ray detected $z\ge4$ AGNs.}
Figure~3 shows the observed-frame, Galactic absorption-corrected \hbox{0.5--2~keV} 
flux versus \ab1450\ magnitude for a compilation of $z\ge4$ AGNs 
(most of these are radio-quiet quasars; RQQs) with \hbox{X-ray} detections or upper limits. 
The majority of the objects shown in Fig.~3 are optically selected 
%
\figurenum{3}
\centerline{\includegraphics[width=8.5cm,angle=0]{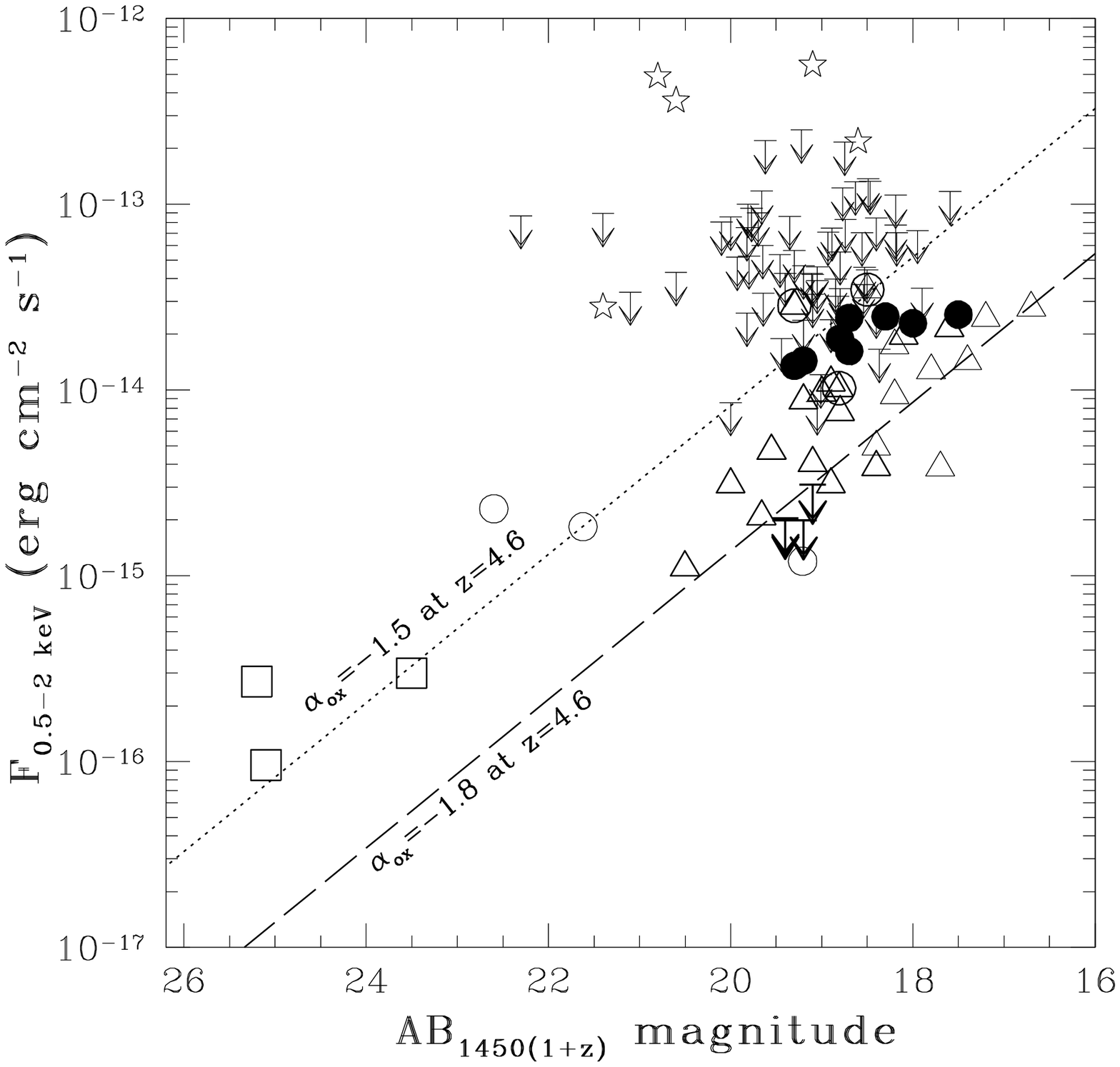}}
\vglue-0.35cm \figcaption{\footnotesize 
Observed-frame, Galactic absorption-corrected 0.5--2~keV flux 
versus \ab1450\ magnitude for $z\ge4$ AGNs. 
The three $z>4$ AGNs in the \cdfn\ are plotted as open squares. 
Open triangles and thick downward-pointing arrows indicate \chandra\ observations of 
$z\ge4$ quasars (V01; V03; Brandt et al. 2002a; Bechtold et al. 2002); 
circled triangles are radio-loud quasars. 
The quasars with \rosat\ detections or upper limits are plotted as 
filled circles and thin downward-pointing arrows, 
respectively (Kaspi et al. 2000; V01; V03); 
blazars are shown as open stars. 
Open circles show X-ray detected AGNs from other 
work (Schneider et al. 1998; Brandt et al. 2001a; Silverman et al. 2002). 
For comparison, the slanted lines show $z=4.6$ loci for \aox=$-$1.5 (dotted) and \aox=$-$1.8 
(dashed).
\label{fig3}}
\vglue0.2cm
%
\noindent 
and populate the bright (upper right) region of the plot. 
By contrast, the sources under investigation in this Letter (open squares) populate the faint (lower left) 
region. At present, 
the study of the properties of such faint sources via \hbox{X-ray} spectral analysis 
is possible only with deep \chandra\ and \xmm\ exposures. 

Figure~3 shows that the optically selected quasars have, on average, steeper \aox\ than 
the \hbox{X-ray} selected AGNs in our sample, as expected because of the different selection criteria and the 
anti-correlation found between \aox\ and rest-frame 2500~\AA\ luminosity (e.g., Vignali, Brandt, \& Schneider 2003). 
Using {\sc ASURV} (LaValley, Isobe, \& Feigelson 1992) 
to take into account the \hbox{X-ray} upper limits, we derive $\langle\alpha_{\rm ox}\rangle$=$-$1.74$\pm{0.02}$ 
(the errors represent the standard deviation of the mean) for the sample of $z\ge4$ optically selected RQQs 
observed by \chandra\ to date (see V03). 
The sample of three $z>4$ AGNs in the \cdfn\ have 
$\langle\alpha_{\rm ox}\rangle$=$-$1.48$\pm{0.03}$. 
Excluding \threer\ because of its radio selection and including the 
X-ray selected quasars in the Lockman Hole (Schneider et al. 1998; $z=4.45$) and ChaMP survey (Silverman et al. 2002; $z=4.93$), 
we find $\langle\alpha_{\rm ox}\rangle$=$-$1.42$\pm{0.04}$.

\acknowledgments

This work would not have been possible without the support of the 
entire \chandra\ and ACIS teams.
We particularly thank P.~Broos and 
L.~Townsley for data analysis software and CTI correction support, 
and G.~Brunetti for useful discussions. 
We acknowledge the financial support of NASA grants 
NAS~8-38252 and NAS~8-01128 (GPG, PI), NSF CAREER award AST-9983783 
(CV, FEB, DMA, WNB), CXC grant G02-3187A (FEB, DMA, WNB), 
CXO Grant PF2-30021 (AEH), and NSF grant AST-9900703~(DPS). 
CV also acknowledges partial support from ASI I/R/113/01 and Cofin-00-02-36.

\clearpage

\end{document}

%% file: tabb1.tex
\begin{table*}[t]
\scriptsize
\caption{Properties of $z>4$ AGNs in the \cdfn}
\begin{center}
\begin{tabular}{ccccccccccc}
\hline
\hline
  &  & $I-$band & \ab1450$^{\rm a}$ & & \multicolumn{2}{c}{Counts$^{\rm b}$} & 
0.5--8~keV Effective & \multicolumn{2}{c}{Count~rate$^{\rm c}$} & Off-axis \\
Object Name & $z$ & mag & mag & $M_{\rm B}$ & (0.5--2~keV) & (0.5--8~keV) & Exposure (ks) &
(0.5--2~keV) & (0.5--8~keV) & angle (\arcmin) \\
\hline
\three & 4.424 & 24.9$^{\rm d}$ & 25.1 & $-$21.6 & 31.0$^{+7.3}_{-6.1}$ & 44.6$^{+8.9}_{-7.7}$ & 
1929.5 & 1.61$^{+0.38}_{-0.32}$ & 2.31$^{+0.46}_{-0.40}$ & 0.6 \\
\one   & 5.186 & 23.1$^{\rm e}$ & 23.5 & $-$23.4 & 96.7$^{+11.7}_{-10.5}$ & 137.9$^{+14.6}_{-13.3}$ & 
1814.0 & 5.34$^{+0.64}_{-0.58}$ & 7.60$^{+0.81}_{-0.73}$ & 4.3 \\
\two   & 4.137 & 25.0$^{\rm e}$ & 25.2 & $-$21.4 & 83.1$^{+11.3}_{-10.1}$ & 117.4$^{+14.2}_{-13.0}$ & 
1769.8 & 4.70$^{+0.64}_{-0.57}$ & 6.63$^{+0.81}_{-0.73}$ & 5.2 \\
\hline
\end{tabular}
\vskip 2pt
\parbox{0.99\textwidth}
{\small\baselineskip 9pt
\footnotesize
\indent
%
%
$^{\rm a}$ \ab1450\ magnitudes have been derived from the $I-$band magnitudes 
using extrapolations, i.e., an ultraviolet/optical slope of $\alpha$=$-$0.5 
\hbox{($S_{\nu}$ $\propto$ $\nu^{\alpha}$)}. 
$^{\rm b}$ Source counts and 1$\sigma$ statistical errors (from Gehrels 1986) 
have been calculated using circular aperture photometry; these values have been 
taken from the 2~Ms X-ray catalog (D.~M. Alexander et al., in preparation). 
The number of 0.5--8~keV counts reported here may be different from that used in the spectral analysis 
because of the different methods adopted in their computation; 
see F.~E. Bauer et al., in preparation, for further discussion. 
$^{\rm c}$ Observed count rate (corrected for vignetting using the exposure maps), 
in units of $10^{-5}$~counts~s$^{-1}$. 
$^{\rm d}$ $I-$band magnitude from W99. 
$^{\rm e}$ $I-$band magnitude from B02. 
}
\end{center}
\vglue-0.85cm
\label{tab1}
\end{table*}
\normalsize
%

%% file: tabb2.tex
\begin{table*}[!t]
\scriptsize
\caption{X-ray Spectral Results}
\begin{center}
\begin{tabular}{lccccccccc}
\hline
\hline
  &  & \multicolumn{3}{c}{PL+ABS} & & \multicolumn{2}{c}{Flux$^{\rm a}$} & & \\
\cline{3-5} \cline{7-8} \\
Object Name & PL -- $\Gamma$ & 
$\Gamma$ & $N_{\rm H,z}$~(cm$^{-2}$) & $N_{\rm H,z}^{\, \rm b}$~(cm$^{-2}$) & Fe~K$\alpha$ EW$^{\rm c}$ (eV) & 
(0.5--2~keV) & (0.5--8~keV) & $\log (L_{\rm 2-10~keV})^{\rm d}$  & \aox$^{\rm e}$ \\
\hline
\threer & 1.55$^{+0.61}_{-0.54}$ & 1.55$^{+0.61}_{-0.54}$ & $<4.82\times10^{22}$ & 
$<6.85\times10^{22}$ & $<400$ & 0.9 & 2.7 & 43.2 & $-$1.52 \\
\oner & 1.81$^{+0.31}_{-0.29}$ & 1.76$^{+0.41}_{-0.25}$ & $<7.47\times10^{22}$ & 
$<9.48\times10^{22}$ & $<700$ & 2.9 & 6.8 & 43.9 & $-$1.51 \\
\twor & 1.12$\pm{0.25}$ & 1.34$^{+0.34}_{-0.33}$ & $<1.64\times10^{23}$ & 
1.90$^{+0.94}_{-0.76}\times10^{23}$ & $<660$ & 2.6 & 11.8 & 43.4 & $-$1.41\\
\hline
\end{tabular}
\vskip 2pt
\parbox{0.97\textwidth}
{\small\baselineskip 9pt
\footnotesize
\indent
{\sc Note. ---} 
``PL'' means a power-law model and Galactic absorption; ``PL$+$ABS'' 
indicates that additional neutral absorption at the source redshift has been used in the spectral fitting. \\
%
$^{\rm a}$~Observed-frame flux derived from the \hbox{X-ray} spectral analysis, in units of $10^{-16}$~\cgs. 
$^{\rm b}$~Intrinsic column density computed assuming $\Gamma=2.0$. 
$^{\rm c}$~Rest-frame. 
$^{\rm d}$~2--10~keV rest-frame luminosity corrected for Galactic absorption, in units of erg~s$^{-1}$. 
$^{\rm e}$~$\alpha_{\rm ox}=\frac{\log(f_{\rm 2~keV}/f_{2500~\mbox{\tiny\AA}})}
{\log(\nu_{\rm 2~keV}/\nu_{2500~\mbox{\tiny\AA}})}$, 
where $f_{\rm 2~keV}$ and $f_{2500~\mbox{\tiny\AA}}$ are the flux densities at rest-frame 2~keV and 2500~\AA. 
The monochromatic flux densities have been derived from the $I-$band magnitudes (using $\alpha$=$-$0.5) 
and the observed-frame, Galactic absorption-corrected 0.5--2~keV flux (using the photon index reported in the second column). 
}
\end{center}
\vglue-0.85cm
\label{tab2}
\end{table*}
\normalsize